\documentclass[aps,prb,preprint,groupedaddress,showpacs]{revtex4-1}

\usepackage{indentfirst}
\usepackage{comment}
\usepackage{amsmath}
\usepackage{textcomp}
\usepackage{amssymb}
\usepackage{nicefrac}
\usepackage{marvosym}
\usepackage{yfonts}
\usepackage{svrsymbols}
\usepackage{notes2bib}
\usepackage{yfonts}
\usepackage{courier}
\usepackage{array}
\usepackage{graphicx}
\usepackage{nopageno}
\usepackage{txfonts}
\usepackage[usenames]{color}

\begin{document}

\title{Exact and efficient calculation of derivatives of Lagrange multipliers for molecular dynamic simulations of biological molecules}

\author{Pablo Garc\'ia Risue\~no}
\email[]{garcia.risueno@gmail.com}
\affiliation{Instituto de Biocomputaci\'on y F\'isica de Sistemas Complejos, C/ Mariano Esquillor (Edificio I+D), 50018 Zaragoza, Spain}

\date{\today}

\indent{ }

\begin{abstract}
In the simulation of biological molecules, it is customary to impose constraints on the fastest degrees of freedom to increase the time step. The evaluation of the involved constraint forces must be performed in an efficient manner, for otherwise it would be a bottleneck in the calculations; for this reason, linearly-scaling calculation methods have become widely used. If integrators of order higher than 2 (e.g. Gear predictor-corrector methods) are used to find the trajectories of atoms, the \emph{derivatives} of the forces on atoms with respect to the time also need to be calculated, which includes the derivatives of constraint forces. In this letter we prove that such calculation can be analytically performed with linearly scaling numerical complexity ($\mathcal{O}(N_c)$, being $N_c$ the number of constraints). This ensures the feasibility of constrained molecular dynamics calculations with high-order integrators. 
\end{abstract}


\maketitle

Molecular dynamics (MD) \cite{Jen1998Book,All2004Book} has a central role in computational physics and chemistry. It is specially relevant for the analysis of biological molecules \cite{dna,1751-8121-50-7-074005}, which have wide present and expected applications as medicines and catalysts \cite{Ech2007CoP}. Force fields \cite{Gromos2005jcc,Van2009JCC,Phi2005JCC} are among the favourite methods for the calculation of the interactions which give rise to time evolution of biological systems. When used in MD simulations, these are frequently carried out by imposing \emph{constraints}; this way to proceed makes it possible to remove the fastest degrees of freedom of the system and hence to increase the time steps of the simulation \cite{Eastman083055,Maz1998JPC}, thus reaching higher total simulated times and making it possible to analyse a wider range of phenomena. 
The imposition of constraints demands to modify the equations of motion by including constraint forces. The calculation of these forces must be performed in an efficient manner, for otherwise it may become a bottleneck in the execution. To this end, several linearly scaling constraint solvers\cite{Ryc1977JCoP,Hes1997JCC,And1983JCOP} have acquired great popularity. 
In MD simulations, atomic positions are customarily calculated using integrators of order 2 (i.e., numerical methods which use the first and second derivatives of the positions with respect of the time).
The most popular are Verlet's \cite{Ver1967PR,Lei1994JCoP} and its equivalent Leap-frog \cite{Jen1998Book} algorithms. The most widely used constraint solvers are devised for such schemes.
Nonetheless, higher order integrators, like Gear predictor-corrector algorithms \cite{All2005Book,Cra2002Book}, can be used for higher accuracy of trajectories \cite{Leimkuhlerbook,Jen1998Book}. The employment of algorithms of order higher than 2 can require the calculation of time derivatives of the forces on atoms. For the \emph{external forces} (those not due to constraints), these derivatives can be obtained by finite-difference. The time derivatives of constraint forces must also be calculated in an efficient manner. In this letter we prove that this task can be performed in an analytical manner with linearly scaling numerical complexity.
Analytical methods have been applied for the calculation of constraint forces within the Verlet integration scheme \cite{Bar1995JCC,Bai2008JCoP,Maz2007JPAMT}. Here we extend these methods to calculate their time derivatives as well.

If holonomic, rheonomous constraints are imposed on a classical system of $N_{\atom}$
atoms\cite{svrsymbols}, and the D'Alembert's principle is assumed to hold, the atomic positions are given by
\cite{GR2011JCC}:
\begin{subequations}
\label{sistemBasico}
\begin{align}
m_\alpha \frac{\mathrm{d}^{2}\vec{x}_{\alpha}(t)}{\mathrm{d}t^{2}}
 & = \vec{F}_{\alpha}( \mathbf{x}(t))+
 \sum_{I=1}^{N_{c}}{\lambda_{I}(t)\vec{\nabla}_{\alpha}
  \sigma^{I}(\mathbf{x}(t))} \ ,
  \qquad \alpha=1,\ldots,N_{\atom} \ , \label{newton} \\
\sigma^{I}(\mathbf{x}(t)) & =0 \ , \qquad I=1,\ldots,{N_{c}} \ , 
  \label{constr} \\
\mathbf{x}(t_{0}) & = \mathbf{x}_{0}\ , \label{sistemaBasico_ci1} \\
\frac{\mathrm{d}\mathbf{x}(t_{0})}{\mathrm{d}t} & =\dot{\mathbf{x}}_{0} \label{sistemaBasico_ci2} \ ,
\end{align}
\end{subequations}
where $\alpha$ is the atom index\cite{notacion}, $m_{\alpha}$ is the atom mass, and $\vec{x}_{\alpha}$ is the Euclidean atom position (with $\mathbf{x}$ representing the set of all them). 
Eq. (\ref{constr}) are the constraint equations. Eq. (\ref{newton}) corresponds to the second law of Newton; 
the first term on its right hand side represents the external forces, and
the second term on its right hand side corresponds to the \emph{forces of constraint}.
$\lambda_I$ is the Lagrange multiplier associated to the $I$-th constraint.
The imposition of $N_{c}$ constraints turns the original system of $3N_{\atom}$ 
equations and unknowns into a system of $3N_{\atom}+{N_{c}}$ equations, being (\ref{constr}) the new equations and $\lambda_{I}$, the Lagrange multipliers, the new unknowns.

The fact that  (\ref{constr}) must hold for all (continuous) times implies that the time derivatives of the constraint equations 
(${\mathrm{d}\sigma^{I}}/{\mathrm{d}t}=0$ \ ; 
${\mathrm{d}^{2}\sigma^{I}}/{\mathrm{d}t^{2}}  = 0 $ \ ; $\ldots$) 
must also hold. This is:

\begin{equation}\label{derivsttt}
\sigma^{I}(\mathbf{x}(t))=0  \quad \forall t \qquad \Rightarrow  \qquad (\sigma^{I})^{(s)}(\mathbf{x}(t))=0 \qquad \textrm{for $s$} \ \in \mathbb{N}   \quad \forall t  \ ,
\end{equation}
where the $(s)$ superscript indicates $s$-th time derivative. This leads \cite{GR2011JCC} to:

%

\begin{subequations}
\begin{align}
\label{Rl}
\frac{\mathrm{d}^{2}\sigma^{I}}{\mathrm{d}t^{2}}&=
\sum_{\mu} \frac{1}{m_{\mu}} \left(F_{\mu}+
\sum_{J}\lambda_{J}\frac{\partial \sigma^{J}}{\partial 
 x^{\mu}}\right)\frac{\partial \sigma^{I}}{\partial x^{\mu}}+
 \sum_{\mu,\nu}\frac{\mathrm{d} x^{\mu}}{\mathrm{d}t}
 \frac{\mathrm{d} x^{\nu}}{\mathrm{d}t}\frac{\partial^{2}
  \sigma^{I}}{\partial x^{\mu} \partial x^{\nu}}  = o^{I} +\sum_{J}R_{IJ}\lambda_{J}=0  \ ,
\intertext{where}
o^{I} & :=p^{I}+q^{I} \ , \qquad I=1,\ldots, N_{c} \ ,  \label{defO} \\
p^{I} & :=\sum_{\mu} 
  \frac{1}{m_{\mu}}F_{\mu}\frac{\partial \sigma^{I}}{\partial x^{\mu}}=
  \sum_{\alpha} \frac{1}{m_{\alpha}}
  \vec{F}_{\alpha} \cdot \vec{\nabla}_{\alpha}\sigma^{I} \ ,
  \label{PQ1} \\
q^{I} & := \sum_{\mu,\nu}\frac{\mathrm{d} x^{\mu}}{\mathrm{d}t}\frac{\mathrm{d}x^{\nu}}{\mathrm{d}t}
  \frac{\partial^{2} \sigma^{I}}{\partial x^{\mu} \partial x^{\nu}} \ ,
  \label{PQ2} \\
R_{IJ} & := \sum_{\mu} \frac{1}{m_{\mu}}\frac{\partial \sigma^{I}}{\partial
  x^{\mu}}\frac{\partial \sigma^{J}}{\partial x^{\mu}}= 
  \sum_{\alpha}{\frac{1}{m_{\alpha}}\vec{\nabla}_{\alpha}\sigma^{I}\cdot
  \vec{\nabla}_{\alpha}\sigma^{J}} \ . \label{defR}
\end{align}
\end{subequations}
Equations (\ref{Rl}) make it possible to calculate the Lagrange multipliers, and hence the forces of constraint, 
as a function of known (unconstrained) atomic positions and velocities.
In Ref. [\onlinecite{GR2011JCC}] it was proved that in biological molecules it is possible to analytically calculate them with linear scaling, i.e. involving $\mathcal{O}(N_{c})$ floating point operations (which, in molecules consisting of up to thousands of atoms, is expected to be performed very efficiently, given the present performance of computing facilities \cite{reviewibanez}).

%
%

In the seminal Reference [\onlinecite{Ryc1977JCoP}]
the following expression is given for the position of atom $\alpha$ in a system subject to
constraints $\sigma^I(x)=0$, $I=1, \ldots, N_c$

\begin{eqnarray}\label{posicshake}
\vec{x}_{\alpha}(t)&=&\vec{x}_{\alpha}(t_{0}) + \left( \frac{d\vec{x}_{\alpha}}{dt}(t_{0})  \right) \cdot(t-t_{0})
 \,   \\
&& \qquad  \, \, \, +\frac{1}{m_{\alpha}}\sum_{l=2}^{\infty} \, \left(\vec{F}_{\alpha}^{(l-2)}(t_{0})-\sum_{I=1}^{N_{c}}\sum_{k=0}^{l-2}\binom{l-2}{k}
\lambda_{I}^{(k)}(t_{0})(\vec{\nabla}_{\alpha}\sigma^{I})^{(l-2-k)}(t_{0})\right) \, \cdot \left( \frac{(t-t_{0})^{l}}{l!}\right)  \ ,\nonumber
\end{eqnarray}
where the superscripts in brackets --$(l)$-- represent ($l$-th) time derivatives, 
$\vec{F}_{\alpha}$ is the external force acting on atom $\alpha$ and $\lambda_I$ is the Lagrange 
multiplier associated to the constraint $\sigma^I(x)$. This 
equation is obtained from performing Taylor series for $\vec{x}_{\alpha}$. 
The $\sum_l$ above is customarily truncated at $l=2$. Nonetheless, eq. (\ref{posicshake}) 
states that the MD calculations can require derivatives of Lagrange multipliers up to an arbitrary order.

Eq. (\ref{derivsttt}) implies \cite{Ryc1977JCoP}

\begin{eqnarray}\label{shake1}
(\sigma^{I})^{(s+2)}&&(\vec{x}(t_{0}))=\sum_{\gamma=1}^{N_{\atom}}\sum_{k=0}^{s+1}
\binom{s+1}{k}\vec{\nabla}_{\gamma}(\sigma^{I})^{(k)}(t_{0})\cdot \\
&& \qquad \qquad \qquad \cdot \, \frac{1}{m_{\gamma}} \left(\vec{F}_{\gamma}^{(s-k)}(t_{0})-\sum_{J=1}^{N_{c}}\sum_{l=0}^{s-k}\binom{s-k}{l}
\lambda_{J}^{(l)}(t_{0})\vec{\nabla}_{\gamma}(\sigma^{J})^{(s-k-l)}(t_{0}) \right)=0 \nonumber
\ , 
\end{eqnarray}
which we express as
\begin{subequations}
\begin{align}\label{shake1mod}
\sum_{J=1}^{N_{c}}R_{IJ}\lambda_{J}^{(s)}(t_{0}) & =\mu_{I}^{(s)}+\sum^{\quad \, \, \prime}_{l} 
\Big(\sum_{J=1}^{N_{c}}(R'^{l}\big)_{JI}\Big)\lambda_{J}^{(l)}(t_{0}) \ , \\
\intertext{where}
R_{JI}& := \sum_{\zeta=1}^{N_{\atom}}\frac{1}{m_{\zeta}}\vec{\nabla}_{\zeta}\sigma^{J}(t_{0})\vec{\nabla}_{l}\sigma^{I}(t_{0}) \ ,\label{defsaux1} \\
\mu_{I}^{(s)}& := \sum_{\zeta=1}^{N_{\atom}}\sum_{k=1}^{s+1}\frac{1}{m_{\zeta}}\vec{F}_{\zeta}^{(s-k)}(t_{0})\vec{\nabla}_{\zeta}(\sigma^{I})^{(k)}(t_{0}) \ ,\label{defsaux2}\\
(R'^{l}\big)_{JI}& := \sum_{\zeta=1}^{N_{\atom}}\sum_{k}\binom{s+1}{k}\binom{s-k}{l}  \, \, \frac{1}{m_{\zeta}}
\vec{\nabla}_{\zeta}(\sigma^{J})^{(s-k-l)}(t_{0})\vec{\nabla}_{\zeta}(\sigma^{I})^{(k)}(t_{0}) \ . \label{defsaux3}
\end{align}
\end{subequations}
The prime in the superscript of $\sum_{l}^{\prime}$ means that the term
$k=0$, $l=s$ is to be skipped.  We express eq. (\ref{shake1mod}) in matrix form as:
\begin{equation}\label{shake1modmod}
R\vec{\lambda}^{(s)}=\vec{\mu}+{R'}^{1}\vec{\lambda}^{(1)}+{R'}^{2}\vec{\lambda}^{(2)}+\ldots+{R'}^{s-1}\vec{\lambda}^{(s-1)} \ .
\end{equation}
This equation states that the time derivatives of the Lagrange multipliers of a given order can be obtained from the time derivatives at lower orders; their calculation also requires to know the derivatives of the external forces, which are included in the independent term $\mu$.

For the sake of simplicity, in our derivation we consider the standard form of constraints in MD simulations of biological molecules, i.e. constraints which freeze the distance between pairs of atoms:
\begin{equation}
\label{sigma_generica}
\sigma^{I(\alpha,\beta)}(\mathbf{x})
 :=|\vec{x_{\alpha}}-\vec{x}_{\beta}|^{2}-(a_{\alpha,\beta})^{2}\ ,
\end{equation}
being $a_{\alpha,\beta}$ is a constant.
Note that a given constraint ($I$) involves a given pair of atoms ($\alpha,\beta$).
Eq. (\ref{sigma_generica}) can represent a constraint on:

\begin{itemize}
\item a bond length between atoms $\alpha$ and $\beta$,
\item a bond angle between atoms $\alpha$, $\beta$ and $\gamma$, if both
$\alpha$ and $\beta$ are connected to $\gamma$ through constrained bond
lengths,
\item a principal dihedral angle involving $\alpha$, $\beta$, $\gamma$ and
$\delta$ (see Ref. [\onlinecite{Ech2006JCCa}] for a rigorous definition of the different
types of internal coordinates), if the bond lengths ($\alpha,\beta$),
($\beta,\gamma$) and ($\gamma,\delta$) are constrained, as well as the bond
angles ($\alpha,\beta,\gamma$) and ($\beta,\gamma,\delta$),
\item or a phase dihedral angle involving $\alpha$, $\beta$, $\gamma$ and
$\delta$ if the bond lengths ($\alpha,\beta$), ($\beta,\gamma$) and
($\beta,\delta$) are constrained, as well as the bond angles
($\alpha,\beta,\gamma$) and ($\alpha,\beta,\delta$).
\end{itemize}

Despite the fact that we use eq. (\ref{sigma_generica}) in our derivation, our conclusion (i.e. the linear scaling of the calculation of time derivatives of Lagrange multipliers) is valid for more general expressions for the constraints. This is because a constraint equation of bond lengths, bond angles or dihedral angles in biological molecules can be expressed as a function of a low number of atomic coordinates (compared to the total number of atoms of the molecule), which makes the coordinate matrix $R$ sparse. Moreover, the topology of biological molecules, which is frequently essentially linear\cite{noteSSS}, makes the coordinated matrix ($R$) banded, which makes its solution still more efficient than that of a generic sparse matrix.

From eq.~(\ref{sigma_generica}) it is straightforward that
\begin{equation}
\label{grad_sigma}
\vec{\nabla}_{\gamma}\sigma^{I(\alpha,\beta)}=
2(\vec{x_{\alpha}}-\vec{x}_{\beta})(\delta_{\gamma,\alpha}-
 \delta_{\gamma,\beta}) \ ,
\end{equation}

where $\delta_{\alpha,\beta}$ is the Kronecker delta. Eq. (\ref{grad_sigma}) corresponds to an $R$ matrix (\ref{defsaux1}) given by:

\begin{eqnarray}
\label{sparseR}
R_{I(\alpha,\beta),J(\gamma,\epsilon)} & := & 
 \sum_{\zeta=1}^{N_{\atom}}{\frac{1}{m_{\zeta}}
   \vec{\nabla}_{\zeta}\sigma^{I(\alpha,\beta)} \cdot 
   \vec{\nabla}_{\zeta}\sigma^{J(\gamma,\epsilon)}} \nonumber \\
 & = & \sum_{\zeta=1}^{N_{\atom}}\frac{4}{m_{\zeta}}
      (\vec{x}_{\alpha}-\vec{x}_{\beta}) \cdot 
      (\vec{x}_{\gamma}-\vec{x}_{\epsilon})
      (\delta_{\zeta,\alpha}-\delta_{\zeta,\beta})
      (\delta_{\zeta,\gamma}-\delta_{\zeta,\epsilon}) \nonumber \\
 & = & 4 (\vec{x}_{\alpha}-\vec{x}_{\beta}) \cdot 
         (\vec{x}_{\gamma}-\vec{x}_{\epsilon})
  \left( \frac{\delta_{\alpha,\gamma}}{m_{\alpha}}
        -\frac{\delta_{\alpha,\epsilon}}{m_{\alpha}}
        -\frac{\delta_{\beta,\gamma}}{m_{\beta}}+
        \frac{\delta_{\beta,\epsilon}}{m_{\beta}}\right) \ ,
\end{eqnarray}
where $I$ ($J$) is the constraint which links atoms $\alpha$ and $\beta$ ($\gamma$ and $\epsilon$), and where
we have used that
\begin{equation}
\sum_{\zeta=1}^{N_{\atom}}\delta_{\zeta,\alpha}\delta_{\zeta,\beta}
 =\delta_{\alpha,\beta} \ .
\end{equation}

Operating in the same manner, (\ref{sigma_generica}) implies that  (\ref{defsaux2}) and (\ref{defsaux3}) become

\begin{subequations}\label{refsaux2}
\begin{align}
\mu_{I}^{(s)}&=2\sum_{k=1}^{s+1}\left(\frac{\vec{F}_{\alpha}^{(s-k)}}{m_{\alpha}}-\frac{\vec{F}_{\beta}^{(s-k)}}{m_{\beta}}\right)(\vec{x}_{\alpha}^{(k)}-\vec{x}_{\beta}^{(k)}) \ ,\\
(R'^{l}\big)_{JI}&=\sum_{k=0}^{s+1}\binom{s+1}{k}\binom{s-k}{l}(\vec{x}_{\gamma}^{(s-k-l)}-\vec{x}_{\epsilon}^{(s-k-l)})
\cdot (\vec{x}_{\alpha}^{(k)}-\vec{x}_{\beta}^{(k)}) 
\  \left(\frac{\delta_{\alpha \gamma}}{m_{\alpha}}-\frac{\delta_{\beta \gamma}}{m_{\beta}}-\frac{\delta_{\alpha\epsilon}}{m_{\alpha}}+\frac{\delta_{\beta\epsilon}}{m_{\beta}}\right) \ ,
\end{align}
\end{subequations}
where $\vec{F}_{\alpha}^{(-1)}:= m_{\alpha}d\vec{x}_{\alpha}/dt$.
Therefore one can calculate the components of vector $\vec{\mu}$ using $\mathcal{O}(s+1)$ operations ($s$ is the
order of the derivative of the Lagrange multipliers we are calculating). In addition,
${R}'$ is sparse and it presents the same sparsity pattern (i.e. with 0 in the same entries) as $R$. Hence
the product of ${R'}^{l}$ with a vector of $N_{c}$ components can be obtained in $\mathcal{O}\big(N_{c}\cdot(s+2) \big)$ operations. 
Finally, eq. (\ref{shake1modmod}) guarantees that $\vec{\lambda}^{(s)}$, the sought $N_{c}$-component vector of
the $s$-th derivative of the Lagrange multipliers,
can be obtained in order $N_{c}\cdot (s+2)$ operations. This can be attained by solving the linear system with the procedure presented in Refs. [\onlinecite{GR2011JCC,GR2010JCoP}].
Further gains in efficiency can be reached by following the procedures presented in Ref. [\onlinecite{cckmmd}].

As an example, a possible algorithm of third order (though the concept it is based on can be used to build methods of any order) could be based on:
\begin{align}\label{rPolTay}
\vec{x}_{\alpha}(t_{0}+\Delta t) &=\vec{x}_{\alpha}(t_{0})+\frac{d\vec{x}_{\alpha}(t_{0})}{dt}\Delta
t+\frac{1}{m_{\alpha}}\Big(\vec{F}_{\alpha}(t_{0})+\sum_{I}\lambda_{I}\vec{\nabla}_{\alpha}\sigma^{I}(t_{0})\Big)\Delta t^{2} \nonumber\\
& +\frac{1}{m_{\alpha}}\left(\frac{d\vec{F}_{\alpha}(t_{0})}{dt}+\sum_{I} \, \frac{d\lambda_{I}(t_{0})}{dt} \, \vec{\nabla}_{\alpha}\sigma^{I}(t_{0})+\sum_{I}\lambda_{I}\sum_{\beta}\frac{\partial^{2}\sigma^{I}(t_{0})}{\partial_{\beta}\partial_{\alpha}}\frac{\vec{x}_{\beta}(t_{0})}{dt}\right)\Delta
t^{3} \ 
\end{align}
\noindent{where we have applied the chain rule ${d\sigma^{I}(t_{0})}/{dt}=\sum_{\beta}\vec{\nabla}_{\beta}\sigma^{I}(t_{0})\, \cdot {d\vec{x}_{\beta}(t_{0})}/{dt} $ (in order to ensure that the atomic positions stay within the constrained subspace, one could add one term which compensates the drift due to non-zero time step as performed by LINCS \cite{Hes1997JCC}). 

In conclusion, we have proved that it is possible to analytically calculate the derivatives of constraint forces with respect of time in biological molecules, and to make it with linearly scaling numerical complexity (proportional to the first power of the number of constraints involved). 
This result will enable to perform accurate constrained molecular dynamics simulations using high-order integrators.
}


\begin{thebibliography}{29}
\bibitem [{not({\natexlab{a}})}]{notacion}%
  \BibitemOpen
  \bibinfo {note} {We use the notation which follows for indices:
  \begin{itemize} \item $\alpha,\beta,\gamma,\epsilon,\zeta=1,\ldots,N_{\atom}$
  for atoms. \item $\mu,\nu=1,\ldots,3N_{\atom}$ for atom coordinates. \item
  $I,J=1,\dots,N_c$ for constraints, as well as for rows and columns of the
  associated matrices. \item $k,l$ as generic indices for products and sums.
  \item $s$ for derivative order. \end{itemize} In addition, we use arrows
  (e.g. $\vec{x}$) to represent vectors of 3 components (Cartesian
  coordinates)}\BibitemShut {NoStop}%
\bibitem [{not({\natexlab{b}})}]{noteSSS}%
  \BibitemOpen
  \bibinfo {note} {The words ''essentially linear'' mean that one can number
  the atoms (and the constraints) of a biological molecule so that every one is
  linked only with atoms with a similar index (e.g. atom 100 can be jointed to
  atom 110, but not to atom 1000). This is intuitive because molecules like
  peptides, proteins and nucleic acids form chains, instead of 2D structures
  like fullerenes or 3D structures like solids. Exceptions to the linearity of
  biological molecules are e.g. proteines with S-bonds and triglycerides. Our
  method is also applicable to these cases, though it would be a bit less
  efficient.}\BibitemShut {Stop}%
\end{thebibliography}

%

\end{document}